\documentclass[journal]{IEEEtran}

\usepackage{mathrsfs}
\usepackage{bbm}
\usepackage{amssymb}
\usepackage{bbding}
\usepackage{threeparttable}
\usepackage[mathcal]{euscript}
\usepackage{psfrag,calc,url,bm}
\usepackage{cite}
\usepackage{graphicx}
\usepackage{psfrag}
\usepackage{diagbox}
\usepackage{subfigure}
\usepackage{hyperref}
\usepackage{url}
\usepackage{stfloats}
\usepackage{amsmath}
\usepackage{float}
\usepackage{algorithmic}
\usepackage[ruled,linesnumbered,vlined]{algorithm2e}
\usepackage{color}
\usepackage{boxedminipage}
\usepackage{amsthm}
\usepackage{multirow}
\usepackage{setspace}
\usepackage{soul}
\usepackage{epstopdf}
\newtheorem{Remark}{Remark}
\begin{document}

\title{GNN-Based Beamforming for Sum-Rate Maximization in MU-MISO Networks}
\author{Yuhang Li, Yang Lu,~\IEEEmembership{Member,~IEEE}, Bo Ai,~\IEEEmembership{Fellow,~IEEE}, Octavia A. Dobre,~\IEEEmembership{Fellow,~IEEE}, \\Zhiguo  Ding,~\IEEEmembership{Fellow,~IEEE}, Dusit Niyato,~\IEEEmembership{Fellow,~IEEE}
%\thanks{This work was supported in part by the Fundamental Research Funds for the Central Universities under Grant 2021RC204, in part by National Natural Science Foundation of China (NSFC) under Grant 62101025 and 62221001, in part by China Postdoctoral Science Foundation under Grant BX2021031 and 2021M690342, and in part by Beijing Nova Program under Grant Z211100002121139. \emph{(Corresponding author: Yang Lu.)}}
\thanks{Yuhang Li and Yang Lu are with the School of Computer and Information Technology, and also with the Collaborative Innovation Center of Railway Traffic Safety, Beijing Jiaotong University, Beijing 100044, China (e-mail: 22125206,yanglu@bjtu.edu.cn).}
\thanks{Bo Ai is with the State Key Laboratory of Rail Traffic Control and Safety, and the School of Electronics and Information Engineering, Beijing Jiaotong University, Beijing 100044, China (e-mail: boai@bjtu.edu.cn).}
\thanks{ Octavia A. Dobre is with the Faculty of Engineering and Applied Science, Memorial University, St. John’s, NL A1C 5S7, Canada (E-mail: odobre@mun.ca).}
\thanks{Zhiguo Ding is with Department of Electrical Engineering and Computer Science, Khalifa University, Abu Dhabi, UAE, and Department of Electrical and Electronic Engineering, University of Manchester, Manchester, UK (e-mail: zhiguo.ding@manchester.ac.uk).}
\thanks{Dusit Niyato is with the School of Computer Science and Engineering, Nanyang Technological University, Singapore 639798 (e-mail: dniyato@ntu.edu.sg).}
}

\maketitle
\begin{abstract}
The advantages of graph neural networks (GNNs) in leveraging the graph topology of wireless networks have drawn increasing attentions. This paper studies the GNN-based learning approach for the sum-rate maximization in multiple-user multiple-input single-output (MU-MISO) networks subject to the users' individual data rate requirements and the power budget of the base station. By modeling the MU-MISO network as a graph, a GNN-based architecture named CRGAT is proposed to directly map the channel state information to the beamforming vectors. The attention-enabled aggregation and the residual-assisted combination are adopted to enhance the learning capability and avoid the oversmoothing issue. Furthermore, a novel activation function is proposed for the constraint due to the limited power budget at the base station. The CRGAT is trained in an unsupervised learning manner with two proposed loss functions. An evaluation method is proposed for the learning-based approach, based on which the effectiveness of the proposed CRGAT is validated in comparison with several convex optimization and learning based approaches. Numerical results are provided to reveal the advantages of the CRGAT including the millisecond-level response with limited optimality performance loss, the scalability to different number of users and power budgets, and the adaptability to different system settings.
\end{abstract}

\section{INTRODUCTION}
With the exponential growth of the mobile data demand, enhancing the spectral efficiency (SE) becomes an urgent need for wireless communication systems to accommodate higher transmission data rates and support massive access services \cite{intro1}. As a key SE enabler, the smart antenna technology has drawn great attentions, which makes the beamforming design an important topic \cite{intro2}. Over the past decades, the beamforming design has been mainly solved by  traditional optimization algorithms, such as successive convex approximation (SCA) \cite{SCA1} and block successive upper bound minimization (BSUM) \cite{BSUM1}. However, these high computational complexity due to the iterations makes traditional optimization algorithms hard to realize the real-time processing in practical time-varying wireless communication systems. Besides, the beamforming design problems became extremely complicated for large scale wireless networks \cite{yang_OAC}. 
%Beamforming is a key technique for improving the quality of service in multi-antenna multi-user systems\cite{BF1,BF2}. Over the decades, there has been extensive research into beamforming techniques, such as the sum rate maximization problem\cite{rate1,rate2,rate4}, the energy efficiency maximization problem \cite{EE1,EE2,EE3} and the power minimization problem  \cite{power1,power2,power3,power4} . Traditional optimization algorithms such as successive convex approximation (SCA) and block successive upper bound minimization (BSUM) achieve near-optimal results on these classical problems\cite{SCA1,SCA2,BSUM1}. Despite their effectiveness, the high time complexity of such algorithms can prevent them from meeting the real-time requirements of wireless communication systems. 
%Inspired by the success of deep learning (DL) in computer vision and natural language processing, many researchers have applied deep learning techniques to tackle the challenges faced by existing optimization techniques\cite{DL}. In \cite{SUP,SUP2},  by using supervised learning. The key is that pre-trained neural networks(NN) is used in place of traditional optimisation algorithms to produce near-optimal solutions in a fast, low-complexity, end-to-end manner. However, the performance of supervised learning NN is limited by the WMMSE, which can only produce sub-optimal solutions. 

 Recently, some researchers have developed the deep learning (DL)-based signal processing approaches\cite{DL}, where the traditional optimization algorithms were approximated by neural networks (NNs). In \cite{SUP},  the supervised ``learn-to-optimize'' approach was proposed, where fully connected multi-layer perceptrons (MLPs) were trained to approximate the non-linear mapping between the inputs and outputs of a sum-rate maximization problem. Once the MLPs are well-trained, the average inference time is greatly reduced with limited optimality performance loss to the weighted minimum mean squared error (WMMSE) algorithm. Similarly, in \cite{cnn}, a black-box based convolutional NN (CNN) was trained to approximate the WMMSE algorithm. This supervised learning approach requires a large amount of labeled samples, which burdens the training process. Instead, some works utilized the unsupervised learning to directly optimize the objective function. In \cite{USUP}, the negative objective function was used as the loss function to facilitate an unsupervised learning of NNs to solve a sum-rate maximization problem in a multiple-input single-output (MISO) system.
 %少一个CNN

 The aforementioned models achieve remarkable performance in some scenarios, but the inputs of these models are usually with flat data structures \cite{gnn_data}, which may not be able to exploit the hidden features to improve the learning performance. Due to the graph-structured topology of the wireless networks, the graph NNs (GNNs) become more suitable to approximate the desired transmit design, especially for large-scale or complicated networks \cite{gnn_ben}. Besides, GNNs can be directly generalized to different network scenarios without re-training, which is an important feature in practice, but cannot be realized by the MLPs and CNNs \cite{GNN1}.

There have been some existing works regarding the GNN-based transmit designs, e.g., \cite{CSI_GNN,MP_GNN,li_gat,bip_GNN,im_GNN,het_GNN,he_gnn,GB_gnn,UF_gnn}. 
In \cite{CSI_GNN},  a link scheduling problem was solved by using channel gains and distance information based on graph embedding. In \cite{MP_GNN}, the authors transformed the radio resource management problem into a graph optimization problem and theoretically analysed the substitution equivalence, scalability and generalisation capabilities of GNNs from the perspective of message passing mechanisms. In \cite{li_gat}, a graph attention network (GAT)-based energy efficient beamforming design was proposed  for a multiple-user (MU)-MISO system with the users' individual power constraints. In addition to the fully connected graph representation, the heterogeneous graph representation was studied. For example, in \cite{bip_GNN}, the MU-MISO system was modelled as a bipartite graph, with which, the sum-rate and min-rate maximization problems were solved by a bipartite GNN based on bipartite message passing mechanism. In \cite{im_GNN}, the authors designed a unicast-multicast GNN (UMGNN) and proposed a UMGNN-based approach to  jointly design the multicast and unicast beamformers based on imperfect channel state information (CSI). In \cite{het_GNN}, a device-to-device (D2D) network was modeled as a heterogeneous graph, and the power control and beamforming designs were proposed by use of the heterogeneous interfering GNNs. Nevertheless, only simple constraints were considered in the above works, and thus, the approaches cannot be directly extended to handle the transmit design problems with complicated constraints involving coupling variables, such as the quality-of-service (QoS) constraints.

As the QoS is the central consideration in wireless services, some recent works have studied to QoS-constrained learning approaches. For instance, in \cite{he_gnn}, a user scheduling and beamforming design based on GNN was proposed for MU-MISO downlink systems under constraints of the QoS requirement and the power budget. In \cite{GB_gnn}, a GNN-based approach for primal dual learning was proposed for solving the radio resource management problem with beam selection and link activation constraints for D2D networks. However, the mapping from  CSI to  beamforming vectors was simplified via the WMMSE method, with which only the power parts of the beamforming vectors were the outputs. In \cite{UF_gnn}, the authors proposed a unified framework for GNNS to solve constrained optimization problems in wireless communication, but a proper-selected initial point was required as input. In addition to the above model-based learning approaches, developing the end-to-end learning approach is a fundamental problem for the GNN-based beamforming design. 

%1）有些问题考虑简单的constraint或者没有constraint，但是所研究的问题本身是存在constraint，包括QoS和power budget

%2）模型设计上: a) beam b) 特征使用 c) oversmoothing

%3）如何评价

% However, all of the above work is based on simple constraint problems and focuses only on the design of the model while ignoring solutions to complex constraints, which undoubtedly hinders the practical application of deep learning-based methods. This is an empirical risk minimization under constraints, which is challenging as training must balance optimality and feasibility conditions.

%As mentioned in \cite{GNN1}, designing neural network architecture for GNNs is of high importance to take advantage of GNN-based approaches. GNNs built on fully connected graphs have been proven effective in many scenarios such as MU-MISO and D2D\cite{miso_fg_gnn,d2d_fg_gnn}, but the oversmoothing problem of such GNNs has been ignored. 

The architecture of the GNNs is the key factor to exploit the features \cite{GNN1}. In most aforementioned works, the permutation invariance and permutation equivariance  properties of GNN were leveraged. However, how to further improve the learning capability, such as handling the oversmoothing issue, has yet been investigated. Moreover, as an emerging approach, the evaluation method of the GNN-based approaches is of high importance, but still remains an open issue. The above observations motivate this paper which is to investigate the GNN-based approach to solve the constrained sum-rate maximum problem for MU-MISO networks. The major contributions are summarized as follows:
\begin{itemize}
  \item We formulate a sum-rate maximization problem for MU-MISO networks subject to the rate requirement of each user and the sum-power constraint of the base station (BS). Then, we present a  graph representation of the MU-MISO network, and reformulate the sum-rate maximization problem into a graph optimization problem.     
  \item  To establish the mapping from the CSI to the beamforming vectors, we propose a GNN-based model named CRGAT based on the message passing mechanism. Specifically, we adopt the attention-enabled aggregation and the residual-assisted combination  to enhance the learning capability and avoid the oversmoothing issue. Furthermore, we propose a novel activation function for the sum-power constraint. 
  \item To train the CRGAT in an unsupervised learning manner and reduce the requirement of extensive training data, the rate requirements are incorporated into the objective function by the proposed loss functions based on the penalty method (PM) and Lagrangian duality method (LDM), respectively. With the loss functions, the adaptability of the CRGAT to the considered problem is enhanced.
  \item  We develop an evaluation method for the learning-based approach, based on which we validate the effectiveness of the proposed CRGAT in comparison with several convex optimization and learning based benchmarking schemes. In particular, the millisecond-level response with limited performance loss, the scalability to different numbers of users and power budgets and the adaptability to different system settings are observed for the CRGAT. Besides, the effectiveness of the residual-assisted combination is validated by the ablation experiment, and the transfer learning performance of the CRGAT is illustrated.  
\end{itemize}

{The rest of this paper is organized as follows. Section II introduces the system model and problem formulation. The proposed CRGAT and the unsupervised learning algorithms are respectively presented in Section III and Section IV. Numerical results are presented in Sections V. Finally, Section VI concludes this paper.}

% \textcolor{red}{\emph{Notations}: The following mathematical notations and symbols are used throughout this paper. Boldface lowercase and uppercase letters denote vectors and matrices, respectively.  The sets of n-by-m complex matrices and complex number are denoted by ${\mathbb{C}^{n \times m}}$ and ${\mathbb{C}}$, respectively.  For a complex number $a$, $\left| a \right|$ denotes the its modulus. ${{\rm Re}\{a\}}$  and  ${{\rm Im}\{a\}}$  denote its real and imaginary part, respectively. For a vector $\bf a$, ${\left\| \bf a \right\|_2}$ and ${\| \bf a \|_{\inf}}$  denote the Euclidean norm and the infinity norm, respectively. ${\rm diag}\{\bf a\}$ is a diagonal matrix with the entries of $\{\bf a\}$ on its main diagonal. $\ln(\cdot )$ and ${\rm Prob}( \cdot )$ denote the natural log function and the probability function, respectively. For a matrix ${\bf A}$, ${\bf A}^H$ denotes the conjugate transpose. ${\bf A} \succ 0$ and ${\bf A} \succeq 0$ means ${\bf A}$ is  positive definite and positive semidefinite, respectively. The symbol ${{\mathbb E}}\{.\} $ represents the statistical expectation of the argument. $\{{\bf x}_{ij}\}$ and $\{{\bf x}_{ij}\}_i$ denote all admissible $\{i,j\}$ for ${\bf x}_{ij}$ and all admissible $i$ for ${\bf x}_{ij}$ with fixed $j$, respectively.}

\emph{Notations}: The following mathematical notations and symbols are used throughout this paper. $\bf a$ and $\bf A$ stand for a column vector and a matrix (including multiple-dimensional tensor), respectively. The sets of real numbers and n-by-m real matrices are denoted by ${\mathbb{R}}$ and ${\mathbb{R}^{n \times m}}$, respectively. The sets of complex numbers and $n$-by-$m$ complex matrices are denoted by ${\mathbb{C}}$ and ${\mathbb{C}^{n \times m}}$, respectively. For a complex number $a$, $\left| a \right|$ denotes its modulus. ${{\rm Re}(a)}$  and  ${{\rm Im}(a)}$  denote its real and imaginary part, respectively. For a vector $\bf a$, ${\left\| \bf a \right\|_2}$ is the Euclidean norm. For a matrix ${\bf A}$, ${\bf A}^H$ and $  \left \|{\bf A}\right\|_F$ denote its conjugate transpose and Frobenius norm, respectively. ${\bf A}_{(i,j)}$ and $ {\bf A}_{(i,:)}$  are the $i$-th row and the $j$-th column of matrix $\bf A$ and the $i$-th row of matrix $\bf A$ respectively. For a tensor ${\bf A}$, ${\bf A}_{(i,:,:)}$ denotes the matrix with index $i$ in the first dimension of the tensor.

\section{System Model and Problem Formulation}

\begin{figure}[t]
\begin{center}
{\includegraphics[ width=.5\textwidth]{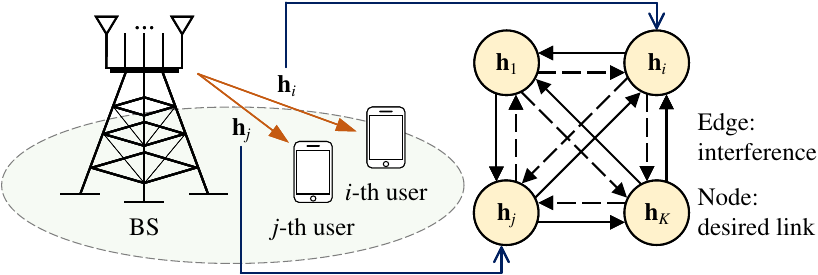}}
\caption{A graph representation of the MU-MISO system.}
\label{sys}
\end{center}
\end{figure}

\subsection{System model}

Consider a downlink MU-MISO system as shown in Fig. \ref{sys}, where one $N_{\rm T}$-antenna BS intends to serve $K$ single-antenna users over a common spectral band. We use $\mathcal{K} \triangleq \{1,2, ..., K\}$ to denote the index set of the users.

Denote the symbol for the $k$-th user and the corresponding beamforming vector by $s_{k}$ and ${{{\bf{w}}_{k}}}$, respectively. The received signal at the $k$-th user is given by
\begin{flalign}
\label{AU_signal}
{{ y}_{k}}&= \underbrace {{\bf{h}}_{k}^H{{\bf{w}}_{k}}{{s}_{k}}}_{\rm desired~signal}+ \underbrace {\sum\nolimits_{i=1,i \ne k}^K {{\bf h}_{k}^H{{\bf w}_{i}}{{s}_{i}}} }_{\mathop{\rm {intra-cell~interference}}} + {n_{k}},\nonumber
\end{flalign}
where ${{{\bf{h}}_{k}}}\in{\mathbb{C}^{{N_{\rm{T}}}}}$ denotes the CSI between the BS and the $k$-th user, and $n_{k}\sim\mathcal{CN}( {0,{\sigma_{k} ^2}})$ denotes the additive white Gaussian noise (AWGN). Without loss of generality, it is assumed that ${{\mathbb E}}\{ {{{| {s_{k} } |}^2}} \} = 1$ ($\forall k\in\mathcal{K}$). Then, the achievable data rate at the $k$-th user is expressed as
\begin{flalign}
{R_{k}}\left( {\left\{ {{{\bf{w}}_{i}}} \right\}} \right) = {\log _2}\left( 1 +\frac{{{{\left| {{\bf{h}}_{k}^H{{\bf{w}}_{k}}} \right|}^2}}}{{{ \sum\nolimits_{i=1,i\ne k}^K {{\left| {{\bf{h}}_{k}^H{{\bf{w}}_{i}}} \right|}^2} + \sigma _{{k}}^2}}} \right).
\end{flalign}

\subsection{Optimization problem formulation}

The sum-rate maximization problem is formulated as
\begin{subequations}\label{pa}
\begin{align}
&\mathop {\max }\limits_{\left\{{{{\bf{w}}_{i}}} \right\}} { {\sum\nolimits_{k = 1}^K {R_{k}}\left( {\left\{ {{{\bf{w}}_{i}}} \right\}} \right)  } }  \label{cons:pa:A}\\
{\rm s.t.}~&{ {\sum\nolimits_{k = 1}^K {\left\| {{{\bf{w}}_{k}}} \right\|_2^2} } } \le {P_{\rm Max}},\label{cons:pa:B}\\
&{R_{k}}\left( {\left\{ {{{\bf{w}}_{i}}} \right\}} \right) \ge {R_{\rm Req}},\label{cons:pa:C}\\
&{{\bf{w}}_{i}}\in{\mathbb{C}^{{N_{\rm{T}}}}},{\forall i,k}\in{\cal K},\nonumber
\end{align}
\end{subequations}
where ${R_{\rm Req}}$ denotes the information rate requirement of each user and ${P_{\rm Max}}$ denotes the power budget of the BS. 

The problem (\ref{pa}) can be solved by conventional convex optimization, which however, requires to be implemented for every CSI realization. To extract the knowledge of the mapping from the CSI to the optimal beamforming vectors, the GNN-based approach is investigated by solving the following graph optimization problem.

\subsection{Graph optimization problem formulation}

%By modeling the considered MU-MISO system as a graph denoted by ${\mathbb G} = ({\mathbb V},{\mathbb E})$, where ${\mathbb V}$ denotes the set of nodes and ${\mathbb E}$ denotes the set of edges. In particular, 

%\begin{figure}[t]
%\begin{center}
%{\includegraphics[ width=.45\textwidth]{representation.pdf}}
%\caption{Graphical representation of the considered MU-MISO network.}
%\label{graph_r}
%\end{center}
%\end{figure}

The network topology nature of the considered MU-MISO system makes it suitable to be modeled as a graph. In particular, the considered MU-MISO system can be modeled as a fully connected directed graph denoted as ${\cal G}=({\cal V},{\cal E})$ as shown in Fig. \ref{sys}, where $\cal V$ denotes the set of nodes and  $\cal E$ denotes the set of edges. Each BS-user desired link is modeled as a node while each interference link is modeled as an edge. That is, there are $K$ nodes and $K(K-1)$ edges. The node feature of the $k$-th node is defined as the corresponding CSI, i.e., ${\bf h}_k$, while there is no edge feature as the inter-user interference is implicit.

The node feature matrix of the graph ${\cal G}$ is denoted as $\mathbf{H}\triangleq [{\bf{h}}_{1}, \cdots, {\bf{h}}_{k} ]^T \in{\mathbb{C}^{{K}\times{N_{\rm{T}}}}}$. By defining the beamforming matrix $\mathbf{W}\triangleq [{\bf{w}}_{1}, \cdots, {\bf{w}}_{k} ]^T\in{\mathbb{C}^{{K}\times{N_{\rm{T}}}}}$, the problem (\ref{pa}) can be rewritten as the following graph optimization problem:
\begin{subequations}\label{pa2}
\begin{align}
&\mathop {\max }\limits_{{{\mathbf{W}}} }~ { {\sum\nolimits_{k = 1}^{K}{{\widehat R_{k}}\left(\mathbf{W}\right)}} }  \label{cons:pa2:A}\\
{\rm s.t.}~&{ {\Vert \mathbf{W} \Vert_F^2} } \le {P_{\rm Max}},\label{cons:pa2:B}\\
&{{\widehat R_{k}}\left(\mathbf{W}\right)} \ge {R_{\rm Req}},\label{cons:pa2:C}\\
&{\mathbf{W}}\in{\mathbb{C}^{{K}\times{N_{\rm{T}}}}},{\forall k}\in {\cal K},\nonumber
\end{align}
\end{subequations}
where 
$${{\widehat R_{k}}\left(\mathbf{W}\right)}\triangleq {\log _2}\left( 1 +\frac{{{{\left| {{\mathbf{H}}_{\left(k,:\right)}^H{{\mathbf{W}}_{\left(k,:\right)}}} \right|}^2}}}{{{ \sum\nolimits_{i \in \mathcal{N}\left(k\right)} {{\left| {\mathbf{H}}_{\left(k,:\right)}^H{{\mathbf{W}}_{\left(i,:\right)}} \right|}^2} + \sigma _{{k}}^2}}} \right)$$
and $\mathcal{N}(k)\triangleq {\cal K}/k$ denotes the set of neighbor nodes of the $k$-th node. 

%To solve the problem (\ref{pa2}), the architecture of the proposed GNN is first introduced with a novel activation function to handle the sum-power constraint (\ref{cons:pa:B}) and then, two kinds of loss functions are introduced to facilitate the unsupervised training. 

\begin{figure*}[t]
\begin{center}
{\includegraphics[ width=.8\textwidth]{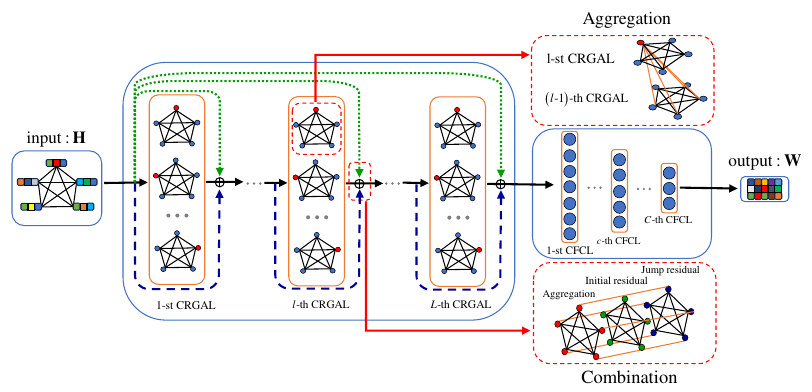}}
\caption{The architecture of the CRGAT.}
\label{GATNN}
\end{center}
\end{figure*}

\section{GNN Architecture}

To solve the defined graph optimization problem, we propose a GNN-based architecture named CRGAT with the node feature matrix (i.e., $\bf H$) as the input and the optimization variable (i.e., $\bf W$) as the output. In particular, the proposed CRGAT is composed of two parts, i.e., a complex encoder and a complex decoder. The complex encoder aims to explore the hidden edge features (such as inter-user interference) as well as exploiting the node features (i.e., $\bf H$) by using the message passing mechanism to obtain embedded features with full graph information. The complex decoder aims to map the embedded features to the learned beamforming matrix. 

The overall architecture of the proposed CRGAT is shown in Fig. \ref{GATNN}, where the complex encoder includes $L$ complex residual graph attention layers (CRGALs) and the complex decoder includes $C$ complex fully connected layers (CFCLs). The details of each layer is presented as follows. 

%In particular, the proposed GNN can be divided into a complex encoder and a complex decoder, and the encoder and decoder contain $L$ complex residual graph attention layers (CRGALs) and $C$ complex fully connected layers (CFCLs), respectively. In addition, ${\mathbb{C} \rm {}SELU}()$ is adopted as the activation function for each layer except for last output layer which does not require an activation function. Besides, a $\mathbb{C} \rm {BatchNorm}$ (CBN) layer is added after each CFCL, which not only prevents the model from overfitting but also enhances the convergence behavior. The overall structure of the proposed GNN is shown in Fig. 2.

\subsection{Complex residual graph attention layer}

The CRGAL is the key component of the proposed CRGAT, and it updates the feature of each node  by aggregating the features of its neighboring nodes. As there is no edge feature, each node can only learn its interference to/from other nodes via the node features. To facilitate the aggregating process, the multiple attention mechanisms are adopted to assign different weights to neighboring nodes\footnote{As the considered system is modelled as a fully connected graph, the neighboring nodes of a node are actually the other $(K-1)$ nodes.}, such that the central node pays more attention to its neighboring nodes with greater impact on it. 

It is noted that the deeper CRGALs may induce the oversmoothing issue, which homogenizes the aggregated features of different nodes and degrades the availability of the aggregated features. Therefore,  the residual structure is adopted to balance the trade-off between adopting deeper CRGALs and avoiding the oversmoothing issue.
\begin{figure}[t]
\begin{center}
{\includegraphics[ width=.48\textwidth]{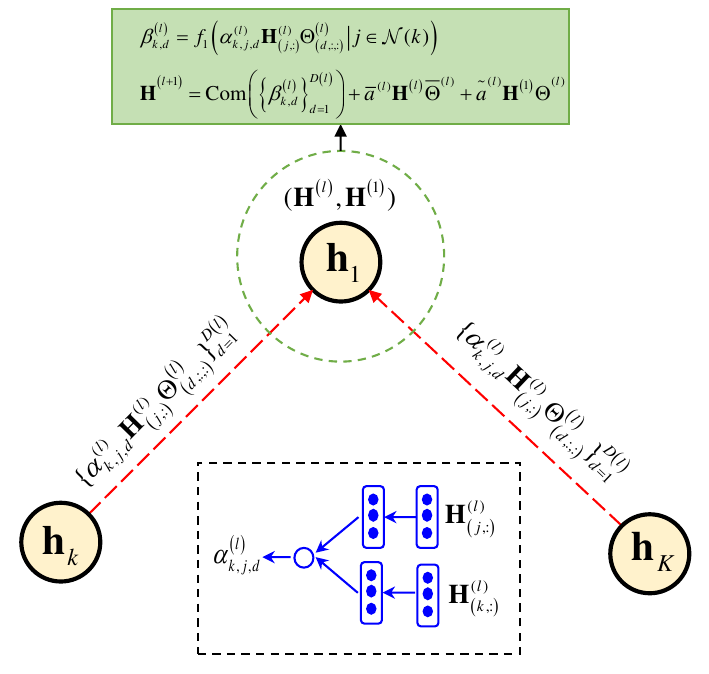}}
\caption{The processes of the CRGAL.}
\label{layer}
\end{center}
\end{figure}

As shown in Fig. \ref{layer}, each CRGAL consists of two processes, i.e., the attention-enabled aggregation and the residual-assisted combination.

\subsubsection{Attention-enabled aggregation}
The $k$-th node aggregates the features of its neighboring nodes using feedforward NNs and attention mechanisms. 

For the $l$-th CRGAL, let $\mathbf{H}^{(l)}\in{\mathbb{C}^{{K}\times{F(l)}}}$ denotes input node feature matrix\footnote{The input node feature matrix of the $1$-st CRGAL is, in fact, the node feature matrix of the graph $\cal G$, i.e., $\mathbf{H}^{(1)}=\mathbf{H}$, and $F(1)=N_{\rm T}$. Besides, the input of the $l$-th CRGAL, i.e., $\mathbf{H}^{(l)}$, is also the output of the $(l-1)$-th CRGAL.} of the $l$-th CRGAL ($l\in {\cal L} \triangleq \{1,...,L\}$), where ${F}(l)$ denotes the corresponding dimension of the node feature. $D(l)$ self-attention mechanisms are applied, and each one is realized by a trainable vector ${\bf a}^{(l)}_{d}\in \mathbb{C}^{{\widetilde F}(l)}$ ($d\in\{1,...,D(l)\} $), where ${\widetilde F}(l)$ denotes the corresponding dimension of the output of the feedforward NN. Denote $\alpha^{(l)}_{k,j,d}$ by the attention coefficient for the $(k,j)$-th node pair after the $d$-th attention mechanism of the $l$-th CRGAL, and $\alpha^{(l)}_{k,j,d}$ is computed by (\ref{attention_co}), where
\begin{flalign}
\mathbf{A}^{(l)} \triangleq [{\bf{a}}^{(l)}_{1}, \cdots, {\bf{a}}^{(l)}_{D(l)} ]^T \in{\mathbb{C}^{{D(l)}\times{\widetilde F}(l)}},
\end{flalign}
${\bm\Theta}^{\left(l\right)} \in {\mathbb{C}^{D(l)\times F(l) \times {\widetilde F}(l)}}$ denotes the learnable parameter of the feedforward NN, $f_2()$ represents a nonlinear operation (e.g., ${\rm LeakyReLU}()$), and ${f}_3():{\mathbb C}\rightarrow{\mathbb R}$ represents a complex-to-real function ({e.g., ${\rm Abs}()$}).

Then, the aggregated feature of the $k$-th node after the $d$-th attention mechanism of the $l$-th CRGAL is computed by 
\begin{flalign}
{\bm \beta}^{\left(l\right)}_{k,d} = f_1\left(\alpha^{(l)}_{k,j,d}{{\mathbf{H}}^{(l)}_{\left(j,:\right)}}{\bm\Theta}_{\left(d,:,:\right)}^{\left(l\right)}  \left|{j \in {\mathcal N}(k)}\right.\right) \in {\mathbb C}^{{\widetilde F}{(l)} },
\end{flalign}
where  $f_1():{\mathbb C}^{{\widetilde F}{(l)}\times |{\mathcal N}(k)| }\rightarrow{\mathbb C}^{{\widetilde F}{(l)} }$ represents a function with permutation invariance (e.g., ${\rm sum}()$). Note that with $D(l)$ independent attention mechanisms, the aggregated features of $D(l)$, i.e., $\{ {\bm \beta}^{(l)}_{k,d} \}_{d=1}^{D(l)}$, are obtained for each node in the $l$-th CRGAL.

\begin{figure*}
\begin{flalign}  
\alpha^{\left(l\right)}_{k,j,d}=\frac{\exp \left({ f}_3\left( {{ f}_2}{\left({\mathbf{H}}^{(l)}_{\left(k,:\right)}{{\bm\Theta}}_{\left(d,:,:\right)}^{\left(l\right)}  + {\mathbf{H}}^{(l)}_{\left(j,:\right)}{{\bm\Theta}}_{\left(d,:,:\right)}^{\left(l\right)} \right)}^T\mathbf{A}^{\left(l\right)}_{\left(d,:\right)}\right)\right)}{\sum\nolimits_{i \in \mathcal{N}\left(k\right)} \exp \left({ f}_3\left({{ f}_2}\left({\mathbf{H}}^{(l)}_{\left(k,:\right)}{{\bm\Theta}}_{\left(d,:,:\right)}^{\left(l\right)}  + {\mathbf{H}}^{(l)}_{\left(i,:\right)}{{\bm\Theta}}_{\left(d,:,:\right)}^{\left(l\right)}\right)^T\mathbf{A}^{\left(l\right)}_{\left(d,:\right)} \right)\right)}\label{attention_co}
\end{flalign}
\hrule
\end{figure*}

\subsubsection{Residual-assisted combination}

The residual-assisted combination is to combine the $D(l)$ attention-enabled aggregated features, the input features of the $l$-th CRGAL and the initial input features (i.e., the input feature of the $1$-st CRGAL). It is noted that the latter two features (also known as the jump residual and the initial residual, respectively) are utilized to overcome the oversmoothing issue. With the residual-assisted combination operation, the net activation of the $l$-th CRGAL denoted as ${\widetilde{\mathbf{H}}}^{(l+1)}$ is given by
\begin{flalign}
{\widetilde{\mathbf{H}}}^{\left(l+1\right)}=& \underbrace {{\rm Com}\left(\left\{ {\bm \beta}^{\left(l\right)}_{k,d} \right\}_{d=1}^{D\left(l\right)}\right)}_{{\rm{Aggregation}}}  + \underbrace {\overline{a}^{(l)} \mathbf{H}^{\left(l\right)}\overline{\bm\Theta}^{(l)}}_{{\rm{
Jump~residual}}} +\\
&\underbrace {\widetilde{a}^{(l)} \mathbf{H}^{\left(1\right)}\widetilde{\bm\Theta}^{(l)}}_{{\rm{Initial~residual}}} \in {\mathbb C}^{K\times ({\widetilde F}(l)\times D(l))},\nonumber
\end{flalign}
where ${\rm Com}()$ represents concatenation operation, $\overline{a}^{(l)}$, $\widetilde{a}^{(l)} \in\mathbb{R}$ are trainable weighting factors. $\overline{\bm\Theta}^{(l)}\in \mathbb{C}^{F(l)\times({\widetilde F}{(l)}\times D(l))}$ and $\widetilde{\bm\Theta}^{(l)}\in \mathbb{C}^{F(1)\times({\widetilde F}{(l)}\times D(l))}$  denote the trainable parameters of the feedforward NNs for the jump residual and the initial residual, respectively.

%&{\rm Com}\left(\left\{ {\bm \beta}^{\left(l\right)}_{k,d} \right\}_{d=1}^{D\left(l\right)} \right) + \overline{a}^{(l)} \mathbf{H}^{\left(l\right)}\overline{\bm\Theta}^{(l)} + \widetilde{a}^{(l)} \mathbf{H}^{\left(1\right)}\widetilde{\bm\Theta}^{(l)},\nonumber

For the $l$-th CRGAL, $\mathbb{C} \rm {SELU}()$ is adopted as the activation function to enhance the representational capability, which is given by 
\begin{flalign}
{{\mathbf{H}}}^{\left(l+1\right)} & = \mathbb{C}\mathrm{SELU}\left({\widetilde{\mathbf{H}}}^{\left(l+1\right)}\right) \\
& =\mathrm{SE}\mathrm{LU}\left({\rm Re} \left({\widetilde{\mathbf{H}}}^{\left(l+1\right)}\right)\right)+i\mathrm{SE}\mathrm{LU}\left({\rm Im}\left({\widetilde{\mathbf{H}}}^{\left(l+1\right)}\right)\right), \nonumber
\end{flalign}
where the activation function ${\rm SELU}()$ is applied to the real and imaginary parts of the input features in an element-wise manner, such that the output and the input of $\mathbb{C} \rm {SELU}()$ have the same shape. Note that the output of the $L$-th CRGAL, i.e., ${{\mathbf{H}}}^{(L+1)}$, is the input of the $1$-st CFCL.

\begin{Remark}\label{Rem1}
The considered system is formulated into a fully connected graph. As a result, each node is able to obtain information from all its neighboring nodes through a single graph convolution, which makes it more likely to suffer from oversmoothing issue. For traditional GAT, the learning performance is degraded by using more layers, because the embedded features of all nodes intend to be identical \cite{ICLR_GAT}. However, the oversmoothing issue is fixed by the adopted residual-assisted combination. As a result, more CRGALs can be stacked while the feature embedding capability is also enhanced. The reason is that the jump residual and the initial residual prevent the  embedded features of all nodes to be identical.
\end{Remark}

\subsection{Complex fully connected layer}

The CFCLs are utilized to  map the embedded features, i.e., ${{\mathbf{H}}}^{\left(L+1\right)}$, to the learned beamforming matrix. Let ${\mathbf{H}}^{\left(c\right)}\in{\mathbb{C}^{{K}\times{G(c)}}}$ denote the input node feature matrix of the $c$-th CFCL ($c\in {\cal C} \triangleq \{1,...,C\}$), where ${{G}(c)}$ denotes the corresponding dimension of each node. In particular, for the $c$-th  CFCL, the mapping from the input to the net activation, denoted by ${\widetilde{\mathbf{H}}^{(c+1)}}$, is realized by
\begin{flalign}
{\widetilde{\mathbf{H}}^{\left(c+1\right)}} = {\mathbf{H}}^{\left(c\right)}{{\bm\Theta}^{(c)}} \in{\mathbb{C}^{{K}\times{G\left(c+1\right)}}},
\end{flalign}
where ${\bm\Theta}^{(c)}\in\mathbb{C}^{G(c)\times G(c+1)}$ denotes the learnable parameters of the $c$-th CFCL. %Note that $G(1) = {{\widetilde F}{(L)}\times D(L)}$.

For the $c$-th CFCL with $c\in {\cal C} \setminus C$, the $\mathbb{C} \rm {BatchNorm}$ ($\mathbb{C}$BN) layer is added in order to prevent the proposed CRGAT from overfitting as well as  enhancing the convergence behavior. The $\mathbb{C}$BN layer can be treated as an activation function, which is applied to the net activation after $\mathbb{C} \rm {SELU}()$ as follows:
\begin{flalign}
{{\mathbf{H}}^{\left(c+1\right)}} =& \mathbb{C}\mathrm{BN}\left(\mathbb{C}\mathrm{SELU}\left({\widetilde{\mathbf{H}}^{\left(c+1\right)}}\right)\right)\\
=&\mathrm{BN}\left(\rm {SELU}\left({\rm Re}\left({\widetilde{\mathbf{H}}^{\left(c+1\right)}}\right)\right)\right)+\nonumber\\
&i\mathrm{BN}\left(\rm {SELU}\left({\rm Im}\left({\widetilde{\mathbf{H}}^{\left(c+1\right)}}\right)\right)\right),\nonumber
\end{flalign}
where ${\rm BN}()$ denotes the batch normalization function. 
%separately in the real and imaginary parts of the input features, defined as follows:

As for the last CFCL, a new activation function is proposed to guarantee that the output beamforming matrix, denoted as $\widehat{ \mathbf{W}}$, satisfies the sum-power constraint (\ref{cons:pa:B}), which is given by
\begin{flalign}\label{af}
&\widehat{ \mathbf{W}} = \phi\left({\widetilde { \mathbf{H}}^{\left(C+1\right)}} \right)\\
&=\left\{\begin{array}{l}
\sqrt {{P_{\rm Max  }}} {{\widetilde { \mathbf{H}}}^{\left(C+1\right)}},~{\left\| {{\widetilde { \mathbf{H}}}^{\left(C+1\right)}}  \right\|_F^2 \le 1} \\
\sqrt {\frac{{{P_{\rm Max }}}}{\left\| {{\widetilde { \mathbf{H}}}^{\left(C+1\right)}}  \right\|_F}} {{{\widetilde { \mathbf{H}}}^{\left(C+1\right)}}},~{\left\| {{\widetilde { \mathbf{H}}}^{\left(C+1\right)}}  \right\|_F^2 > 1}  
\end{array}\right. \in\mathbb{C}^{K\times{N_{\rm{T}}}}.\nonumber
\end{flalign}
Note that $\widehat{\mathbf{W}}$ also represents the output of the CRGAT, i.e., the learned beamforming matrix.

With the proposed activation function (\ref{af}), the problem (\ref{pa}) is equivalently transformed into the following optimization problem:
\begin{subequations}\label{pb}
\begin{align}
&\mathop {\min }\limits_{\bm{\theta}} { {\sum\nolimits_{k = 1}^{K}{-{\widehat R_{k}}\left(\widehat{ \mathbf{W}}|\bm{\theta}\right)}} }  \label{cons:pb:A}\\
{\rm s.t.}~
&{{\widehat R_{k}}\left(\widehat{ \mathbf{W}}|\bm{\theta}\right)} \ge {R_{\rm Req}}, \forall k\in {\cal K},\label{cons:pb:C}
\end{align}
\end{subequations}
where $\bm{\theta}\triangleq\{\mathbf{A}^{(l)},{\bm\Theta}^{(l)},\overline{\bm\Theta}^{(l)},\widetilde{\bm\Theta}^{(l)},{\bm\Theta}^{(c)}\}$ represents all trainable parameters in the proposed CRGAT.

\begin{Remark}\label{Rem3}
The activation function (\ref{af}) is based on the projected gradient descent method, which is utilized to realize the allocation of the power budget. That is, the activation function (\ref{af}) in fact simplifies the task from learning the beamforming vectors to learning the directions of beamforming vectors. Besides, it is observed from (\ref{pb}), the impact of $P_{\rm Max}$ has been integrated into the CRGAT, and thus, the CRGAT should have the generalization capability on $P_{\rm Max}$.
\end{Remark}

\begin{Remark}\label{Rem2}
It is observed from the definition of $\bm \theta$ that all learnable parameters are independent from the number of users. Therefore, the CRGAT still works if the number of users changes. That is, the CRGAT has the generalization capability on the number of users. Intuitively, the generalization performance is affected by the variance of the number of users in training set and test set. Based on the generalization performance, one can directly use the well-trained CRGAT or update the CRGAT in a transfer learning manner, which greatly saves the time and computational resources for re-training a new CRGAT, if the network topology changes.
%\textcolor{green}{In addition the attention mechanism learns the interference relations between channels, while the interference relations between those containing different users are similar, which also ensures good generalization ability of CRGAT.}
\end{Remark}

%\subsection{Complex Scaled Exponential Linear Units}
%To increase the representational capability of the proposed GNN, $\mathbb{C}$SELU is added after all hidden layers, specifically the SELU activation functions are applied to the real and imaginary parts of the input features, respectively, which are given as
%\begin{flalign}
%\mathbb{C}\mathrm{SELU}(z)=\mathrm{SE}\mathrm{LU}(\Re(z))+i\mathrm{SE}\mathrm{LU}(\Im(z)).
%\end{flalign}
%Note that the functions performed in the activation layers are element-wise functions, such that their outputs have the same shapes of their inputs, respectively.

%\subsection{Complex Batch Normalization}
%Deep networks generally rely upon Batch Normalization to accelerate learning. In some cases batch normalization is essential to optimize the model. Batch normalization takes the same treatment as the activation function, separately in the real and imaginary parts of the input features, defined as follows:
%\begin{flalign}
%\mathbb{C}\mathrm{BN}(z)=\mathrm{BN}(\Re(z))+i\mathrm{BN}(\Im(z)).
%\end{flalign}

\section{Unsupervised learning}

With the proposed CRGAT, the mapping from the channel matrix to the learned beamforming matrix satisfying the sum-rate power constraint is established, i.e., $\widehat{\mathbf{W}} = {\rm CRGAT}({\bf H})$. To train the CRGAT via unsupervised learning, the problem (\ref{pa}) should be further transformed into an unconstrained problem, such that the trainable parameters can be trained in a gradient descent manner. 

To this end, the PM and the LDM are respectively adopted to design the loss function based on the objective function (\ref{cons:pb:A}) and the rate requirement constraint (\ref{cons:pb:C}).

\subsection{Loss function based on penalty method}

\begin{figure*}
\begin{flalign}
{{L}_{\rm Penal}\left( {\bm\theta}  \right) = \frac{1}{N} \sum\nolimits_{n=1}^N {\left[\underbrace {\sum\nolimits_{k = 1}^{K}{-{\widehat R_{k}}{\left({\widehat{\mathbf{W}}_{\left(n\right)}}|\bm\theta\right)}}}_{\rm Objective~function~due~to~(\ref{cons:pb:A})}  + \lambda \underbrace {  \sum\nolimits_{k = 1}^{K}{\operatorname{ReLU}\left(R_{\rm Req} - {\widehat R_{k}}\left(\widehat{\mathbf{W}}_{\left(n\right)}|\bm\theta\right)\right)}}_{\rm Penalty~ term~due~to~(\ref{cons:pb:C})}\right]}}\label{penlty_loss}
\end{flalign}
\hrule
\end{figure*}

The PM embeds the rate requirement constraint (\ref{cons:pb:C}) into the loss function as a penalty term to enforce the output of CRGAT to satisfy (\ref{cons:pb:C}).  For the problem (\ref{pb}), the loss function based on the PM is given by (\ref{penlty_loss}), where $N$ denotes the batch size, $\operatorname{ReLU}(x) \triangleq \max (x, 0)$ represents the rectified linear unit activation function, $\widehat{\mathbf{W}}_{\left(n\right)}$ represents the ouput of the CRGAT assosicated with the $n$-th sample, i.e.,  ${\mathbf{H}}_{\left(n\right)}$, and $\lambda > 0$ denotes the penalty coefficient which is treated as a hyperparameter. 

Particularly, the penalty term is imposed only if at least one rate requirement constraint cannot be satisfied. With a larger value of $\lambda$, the rate requirement constraints are more likely to be satisfied, which however, may degrade the impact of the objective function on the loss function. By carefully adjusting the value of $\lambda$, a tradeoff between maximizing the sum rates and satisfying the rate requirement constraints can be achieved. 

The unsupervised training procedure with the loss function based on PM is shown in Algorithm \ref{alg:PM} where $M$, $E$ and $B$ respectively denote the total numbers of samples, epochs and minibatchs in the training dataset, and $\omega$ denotes the learning rate.
\begin{flalign}
\bm{\theta} := \bm{\theta} - \omega \nabla_{\bm{\theta}}{ {L}_{\rm Penal} \left(\bm \theta \right)} \label{update0}
\end{flalign}

% It is not an obvious task to choose penalty coefficients (or Lagrangian multipliers) to minimize constraint violations and balance them with the objective function
\begin{algorithm}[t]
    \caption{Unsupervised Learning with Loss function based on PM}
    \label{alg:PM}
    \renewcommand{\algorithmicrequire}{\textbf{Input:}}
    \renewcommand{\algorithmicensure}{\textbf{Output:}}
    
    \begin{algorithmic}[1]
        \REQUIRE Training dataset $\mathcal{D}=\left\{\mathbf{H}_{\left(n\right)}\right\}_{n=1}^M$\;  %%input
        \ENSURE Learned   $\bm\theta$;    %%output
        % 开始行数
        \STATE  Initialize $\bm\theta$\;
         \FOR{ epoch $e \in [0,1,\dots,E] $}
            \FOR{ index of minibatch $b \in [0,1,\dots,B] $}
                \STATE  Sample the $b$-th minibatch $ \left\{\mathbf{H}_{\left(n\right)}\right\}_{n=1}^N$\;
                \STATE  Calculate the corresponding beamforming matrices $ \left\{{\widehat{\mathbf{W}}}_{\left(n\right)}\right\}_{n=1}^N$ via $\widehat{\mathbf{W}}_{\left(n\right)} = {\rm CRGAT}\left({\bf H}_{\left(n\right)}\right)$, $\forall n$\;
                \STATE Update $\bm{\theta}$ via (\ref{update0})\;
            \ENDFOR       
          \ENDFOR
        \RETURN $\bm\theta$.
    \end{algorithmic}
\end{algorithm}

%\begin{figure}[t]
%\begin{center}
%{\includegraphics[ width=.48\textwidth]{lagrange.pdf}}
%\caption{Lagrangian}
%\label{layer2}
%\end{center}
%\end{figure}

\subsection{Loss function based on Lagrangian duality method}

Instead of using the constant penalty coefficient, the  LDM adopts the learnable Lagrangian multipliers. 

Let $\bm{\mu}\triangleq [\mu_1,...,\mu_K] \in \mathbb{R}_{+}^{K}$ denote non-negative Lagrangian multipliers associated with (\ref{cons:pb:C}). The  Lagrangian dual optimization problem of the problem (\ref{pb}) is given by
\begin{flalign}\nonumber
\max _{\bm{\mu}} \min _{{\bm{\theta}}} { {\cal L} \left(\bm \theta, \bm{\mu} \right)},
\end{flalign}
where ${ {\cal L} \left(\bm \theta, \bm{\mu} \right)}$ is given by  (\ref{Lagrangian}).
\begin{figure*}
\begin{flalign}
{ {\cal L} \left(\bm \theta, \bm{\mu} \right)} = {\sum\nolimits_{k = 1}^{K}{-{\widehat R_{k}}{\left(\widehat{\mathbf{W}}|\bm{\theta}\right)}}}  
+  \sum\nolimits_{k = 1}^{K}{\mu_{k}\operatorname{ReLU}\left(R_{\rm Req} -{\widehat R_{k}}\left({\widehat {\mathbf W}}|\bm{\theta}\right)\right)}\label{Lagrangian}
\end{flalign}
\hrule
\end{figure*}

The Lagrangian dual optimization problem can be solved by alternately optimizing $\bm\theta$ and $\bm\mu$. Similarly, the loss function based on the LDM is given by (\ref{eq1}), and we alternately update $\bm\theta$ and $\bm\mu$ during the training phase. Then, the unsupervised training procedure with the loss function based on LDM is shown in Algorithm \ref{alg:LDM}, where $\tau  > 0$ denotes the step size for updating $\bm{\mu}$.
\begin{flalign}
\bm{\theta} &:= \bm{\theta} - \omega \nabla_{\bm{\theta}}{ {L}_{\rm Lag} \left(\bm \theta, \bm{\mu} \right)} \label{update1}\\
\nabla_{{\mu}_{k}} &:= \nabla_{{\mu}_{k}} + \tau \sum_{n = 1}^{N} \operatorname{ReLU}\left(R_{\rm Req} - { \widehat R_{k}}\left(\widehat{\mathbf{W}}_{(n)}|\bm{\theta}\right)\right)\label{update2}
\end{flalign}

\begin{Remark}
    Different from handling the sum-power constraint via the activation function, both PM and LDM cannot theoretically guarantee a feasible solution. Nevertheless, the efficacy of the two methods has been validated in some existing works numerically, see e.g., \cite{PM1,LDM}. %Although the LDM involves more learnable parameters, it is by no means that the LDM is always superior to the PM, especially when the constraint (\ref{cons:pb:C}) is not tight.     
    %比较PM与LDM
\end{Remark}

%It is not an obvious task to choose penalty coefficients (or Lagrangian multipliers) to minimize constraint violations and balance them with the objective function. Lagrangian duality approach (LDA) utilizing Lagrangian multipliers to solve the constraint problem (9). In addition, unlike the penalty approach where the penalty coefficient is treated as a hyperparameter, the Lagrangian dual method adopts a subgradient approach to determine the suitable Lagrangian multiplier. Let $\bm{\mu} \in \mathbb{R}_{+}^{K}$ be nonnegative dual multipliers associated with constraint (9b). Specifically, problem (9) is transformed into a Lagrangian function as (\ref{Lagrangian}).

%The LDA iteratively optimizes the parameters of the neural network $\bm\theta$ and the Lagrangian multipliers $\bm\mu$ of the pairwise problem during the training process, as shown in Fig \ref{Lagrangian}.  Specifically, when training the parameters of the neural network $\bm\theta$, the dual variables $\bm\mu$ are fixed and the Lagrangian relaxation function (12) is used for the loss function.The loss function is denoted as (\ref{eq1}).
% 需替换违反程度的函数
\begin{figure*}
\begin{flalign}\label{eq1}
{{ {L}_{\rm Lag} \left(\bm \theta, \bm{\mu} \right)} = \frac{1}{N} \sum\nolimits_{n=1}^N {\left[\sum\nolimits_{k = 1}^{K}{-{\widehat R_{k}}{\left({\widehat{\mathbf{W}}_{\left(n\right)}}|\bm\theta\right)}}   + \sum\nolimits_{k = 1}^{K}{\mu_{k}\operatorname{ReLU}\left(R_{\rm Req} -{\widehat R_{k}}\left(\widehat{\mathbf{W}}_{\left(n\right)}|\bm{\theta}\right)\right)}\right]}}
\end{flalign}
\hrule
\end{figure*}
%After updating the parameters of the neural network $\bm\theta$, the subgradient method is employed to update the Lagrangian multipliers using the rules:
%\begin{flalign}
%{\mu_k} = {\mu_k} + \tau \operatorname{ReLU}\left(R_{\rm Req} - 
%{ \widehat R_{k}}\left(\widehat{\mathbf{W}}|\bm{\theta}\right)\right), {\forall k\in K},
%\end{flalign}
%where the hyperparameter $\tau  > 0$ represents the step size for updating $\bm{\mu}$.

\begin{algorithm}[t]
    \caption{Unsupervised Learning with Loss function based on LDM}
    \label{alg:LDM}
    \renewcommand{\algorithmicrequire}{\textbf{Input:}}
    \renewcommand{\algorithmicensure}{\textbf{Output:}}
    
    \begin{algorithmic}[1]
        \REQUIRE Training dataset $\mathcal{D}=\left\{\mathbf{H}_{\left(n\right)}\right\}_{n=1}^M$\;  %%input
        \ENSURE Learned   $\bm\theta$ and $\bm\mu$\;    %%output
        % 开始行数
        \STATE  Initialize $\bm\theta$ and $\bm\mu$\;
        \FOR{ epoch $e \in [0,1,\dots,E] $}
            \STATE {$\nabla_{\bm{\mu}} \triangleq [\nabla_{{\mu}_{1}},...,\nabla_{{\mu}_{K}}] \gets {\bf 0}$}
            \FOR{ index of minibatch $b \in [0,1,\dots,B] $}
                \STATE  Sample the $b$-th minibatch $ \left\{\mathbf{H}_{\left(n\right)}\right\}_{l=1}^N $\;
                \STATE  Calculate the corresponding beamforming matrices $ \left\{{\widehat{\mathbf{W}}}_{\left(n\right)}\right\}_{n=1}^N$ via $\widehat{\mathbf{W}}_{\left(n\right)} = {\rm CRGAT}\left({\bf H}_{\left(n\right)}\right)$, $\forall n$\;
                \STATE Update $\bm{\theta}$ via (\ref{update1})\;
                \STATE Update $\nabla_{{\mu}_{k}}$ via (\ref{update2}), $\forall k\in {\cal K}$\;
            \ENDFOR
            \STATE {$ \bm{\mu} \gets \bm{\mu} + \tau \nabla_{\bm{\mu}}$}\;
        \ENDFOR
        
        \RETURN $\bm\theta$ and $\bm\mu$.
    \end{algorithmic}
\end{algorithm}

\section{Numerical Results}
In this section, numerical results are provided to evaluate the proposed CRGAT, activation function and two loss functions. An evaluation method for the learning-based approach is proposed based on certain basic metrics, which include three aspects: model effectiveness, scalability and adaptability. Besides, the effectiveness of the residual-assisted combination is validate by the ablation experiment, and the transfer learning performance of the CRGAT is illustrated.

%The simulation setup is first described, then a set of metrics is introduced to discuss the the effectiveness of the proposed CRGAL model from a perspective of application, and finally the 

%performance of the CRGAL model is discussed, and finally the robustness of the penalty and Lagrangian dual methods is tested.
\subsection{Simulation setup}
{\bf Simulation scenario:} The number of antennas is set as $N_{\rm T}\in\{8,16\}$. The number of users is set as $K\in\{3,4,5,6,7,8,9\}$. The total power budget is set as $P_{\rm Max} \in \{1,2,3\}$. The rate requirement is set as $R_{\rm Req} \in \{1,2,3\}$ bit/s/Hz. The CSI of each user includes both large-scale path loss and the small-scale fading. The path-loss model is based on $10^{-(140.7+36.7\log_{10}({\widetilde d}))/10}$ where ${\widetilde d}$ (in kilometer) denotes the distance between the BS and the user. Rayleigh fading is adopted for the considered small-scale fading. The noise power spectrum is $162$ dBm/Hz and the bandwidth is $10$ MHz. The average ratio between path loss and noise power is set as $10$ dB. 

{\bf Computer configuration:} The convex formulations and the NNs are respectively processed by convex solver {\tt SeDuMi} under Mathworks MATLAB R2021b and Python 3.8.18 with Pytorch 1.10.0 on a computer with Intel(R) Core(TM) i9-12900K CPU and one NVIDIA RTX 3090 GPU (24 GB of memory).

{\bf Dataset:} {The ratio between the training set, validation set and test set in the dataset is set as $9:1:1$. Since the unsupervised learning is adopted, the training set does not include labels. The labels of the validation set and test set are generated by an SCA-based approach (presented in Appendix A) with the convergence precision of $10^{-4}$. All the datasets prepared for the simulation experiments are shown in Table \ref{table:1}.
\begin{table}[ht]
\centering
\caption{Datasets.}
\begin{tabular}{c|c|c|c|c|c|c}
\hline
No. &$N_{\rm T}$ &$K$ &${P_{\rm Max}}$ &${R_{\rm Req}}$& Size& Type\\
 \hline
 1   &8&3&1&1& 10,000& B\\
 \hline
2   &8&4&1&1& 110,000& A\\
 \hline
 3   &8&5&1&1& 110,000& A\\
  \hline
4   &16&7&1&1& 10,000& B\\
 \hline
5   &16&8&1&1& 110,000& A\\
  \hline
6   &16&9&1&1& 10,000& B\\
 \hline
7   &16&8&1&2& 110,000& A\\
\hline
8   &16&8&1&3& 110,000& A\\
 \hline
9   &16&8&2&1& 110,000& A\\
 \hline
10   &16&8&3&1& 110,000& A\\
 \hline
\end{tabular}
 \begin{tablenotes}
        \footnotesize
        \item {Type A: The training set, validation set and test set are all included.}
        \item {Type B: Only test set is included.}
\end{tablenotes}
\label{table:1}
\end{table}

{\bf GNN architecture:} The detailed architecture of the CRGAT is shown in Table \ref{table:2} where $4$ CGCLs and $3$ CFCLs are included. 
\begin{table}[ht]
\centering
\caption{The architecture of the proposed CRGAT.}
\begin{tabular}{c|c|c|c|c|c|c}
\hline
No. & Type &IFs &OFs &AMs &   $\mathbb{C}$SELU&   $\mathbb{C}$BN\\
 \hline
1 & CRGL  & $N_{\rm T}$&32&10&$\checkmark$& $\times$\\
 \hline
2 & CGCL  & 320&64&10&$\checkmark$& $\times$\\
 \hline 
 3 & CGCL  & 640&128&10&$\checkmark$& $\times$\\
 \hline
 4 & CGCL  & 1280&256&10&$\checkmark$& $\times$\\
 \hline
5 & CFCL  & 2560&1024&-&$\checkmark$&$\checkmark$\\
 \hline
6 & CFCL  & 1024&512&-&$\checkmark$&$\checkmark$\\
 \hline
7 & CFCL  & 512& $N_{\rm T}$ &-& $\times$ & $\times$\\
 \hline
\end{tabular}
 \begin{tablenotes}
        \footnotesize
        \item IFs: The dimension of the input feature, i.e., $F(l)$, $G(c)$.
        \item OFs: The dimension of the output feature.%
        \item AMs: Number of attention mechanisms, i.e., $D(l)$. %
       \item CGCL: Generalized to all complex graph convolution layers.
\end{tablenotes}
\label{table:2}
\end{table}

{\bf Basic Metrics:} The basic metrics are used to evaluate the learning-based approaches on a given test set in terms of the optimality performance, feasible rate and inference time.
\begin{itemize}
  \item [1)] 
  \emph{Optimality performance:} The ratio of the average achievable sum rates (with feasible solutions) by the learning-based approach to the average optimal sum rates. 
  \item [2)] 
  \emph{Feasibility rate:} The percentage of the feasible solutions to the considered problem by the learning-based approach.
  \item [3)]
 \emph{Inference time:} The average running time required to calculate the feasible beamforming vectors with the given CSI by the learning-based approach.  
\end{itemize}

{\bf Advanced Metrics:}  
Based on the basic metrics, we define the following three advanced metrics to evaluate the learning-based approaches. The detailed is described as follows.
\begin{itemize}
  \item [1)]
  \emph{Model effectiveness:} Evaluate the efficacy of the model architecture based on the basic metrics, where the settings of the test set and the training set are identical.\footnote{The identical setting means that the scalar variables are the same while the random variables are i.i.d.}
  \item [2)]
  \emph{Scalability:} Evaluate the efficacy of the learning-based approach on the test set with different settings from the training set. 
  \item [3)]
  \emph{Adaptability:} {Evaluate the efficacy of the activation function and the loss function with different constraints.} %adaptability
\end{itemize}

%In particular, the model efficacy is for the cases that the settings of the test set and the training set are identical\footnote{The identical setting means that the scalar variables are the same while the random variables are i.i.d.}, while the scalability and the adaptability are for the cases that settings of the test set and the training set are different.

\subsection{Model effectiveness}

This subsection shows the performance of the proposed CRGAT based on the basic metrics. Two scenarios are considered, i.e., $(N_{\rm T},K,P_{\rm Max},R_{\rm Req})=(8,4,1,1)$ (Dataset No. 2) and $(N_{\rm T},K,P_{\rm Max},R_{\rm Req})=(16,8,1,1)$ (Dataset No. 5). To fully evaluate the proposed CRGAT, several convex optimization and learning based approaches are adopted as the benchmarks\footnote{The learning-based approaches are according to existing works and extended to solve the considered problem.}.
\begin{itemize}
  \item [1)] 
  {\bf SCA:} The SCA-based algorithm serves as a benchmark to evaluate the optimality and the computational efficiency of the learning-based approaches;       
  \item [2)] 
  {\bf MRT:}  The SCA-based algorithm based on the maximum ratio transmission (MRT) scheme (given in Appendix B);  
  \item [3)]
  {\bf ZF:} The SCA-based algorithm based on the zero-forceing (ZF) scheme (given in Appendix B);  
  \item [4)]
  {\bf CMLP\cite{SUP}:} Direct learning of the mapping from CSI, i.e., $\{{\bf h}_i\}$, to beamforming vectors, i.e., $\{{\bf w}_i\}$ via CFCLs; 
  \item [5)]
  {\bf CGCN\cite{d2d_fg_gnn}:} Topological information among CSI is utilized through graph convolution;
  \item [6)]
  {\bf CGAT\cite{li_gat}:} Additive attention mechanism is used in the process of graph convolution;
  \item [7)]
  {\bf CTGCN\cite{Tran_GNN}:} Key-value attention mechanism is used in the process of graph convolution.
\end{itemize}
  For fair comparison, all learning-based approaches adopt the activation function (\ref{af}) and trained via Algorithm 1. The numerical results are shown in Table \ref{table:5}.

\begin{table}[ht]
\centering
\caption{Model effectiveness.}
\begin{tabular}{c|c|c|c|c|c}
\hline
$N_{\rm T}$ &$K$& Approach & OP & FR & IT\\
 \hline
 \multirow{8}*{8}&\multirow{8}*{4}& SCA &100\%&100\% &1.99s\\
\cline{3-6}
 ~&~& MRT &46.6\%&100\%&1.29s \\
\cline{3-6}
 ~&~& ZF &99.1\%&100\%&1.12s \\
\cline{3-6}
 ~&~& CMLP &79.7\%$^*$&0\%&0.93ms \\
\cline{3-6}
 ~&~& CGCN &84.5\%&99.3\%&0.90ms \\
\cline{3-6}
 ~&~& CGAT &91.0\%&99.6\%&1.66ms \\
\cline{3-6}
 ~&~& CTGCN &85.0\%&99.2\%&2.42ms \\
\cline{3-6}
  ~&~& {\bf CRGAT} &97.1\%&100\%&2.73ms \\
\hline
\multirow{8}*{16}&\multirow{8}*{8}& SCA &100\%&100\%&8.48s \\
\cline{3-6}
 ~&~& MRT &40.6\%&100\%&1.86s \\
\cline{3-6}
 ~&~& ZF &98.9\%&100\%&2.14s \\
\cline{3-6}
 ~&~& CMLP &36.1\%$^*$&0\%&0.94ms \\
\cline{3-6}
 ~&~& CGCN &52.3\%&98.5\%&0.92ms \\
\cline{3-6}
 ~&~& CGAT &84.0\%&99.9\%&1.71ms \\
\cline{3-6}
 ~&~& CTGCN &45.0\%&92.7\%&2.46ms \\
\cline{3-6}
  ~&~& {\bf CRGAT }  &95.4\%&100\%&2.81ms \\
 \hline
\end{tabular}
 \begin{tablenotes}
        \footnotesize
       \item OP: Optimality performance.
       \item FR: Feasibility rate.
       \item IT: Inference time.
       \item $^*$: The optimality performance is calculated with infeasible solutions as there is no feasible solution. 
\end{tablenotes}
\label{table:5}
\end{table}

Compared with the benchmark learning-based approaches, the proposed CRGAT achieves the best optimality performance, and optimality performance gap to the SCA-based algorithm is less than  $5\%$. The increment of antennas and users has an insignificant impact on the optimality performance of the CRGAT, which however, greatly degrades the optimality performance of other learning-based approaches. On the one hand, the attention-enabled aggregation adopted in the CRGAT is able to explore the hidden features due to the inter-user interference, which makes it more suitable to the considered problem than the CMLP and the CGCN. On the other hand, the residual-assisted combination adopted in the CRGAT avoids the oversmoothing issue, which enhances its  feature embedding capability compared with the CGAT and the CTGCN. Besides, the feasible solution by the CRGAT is guaranteed. 
%FR
%increases almost does not affect the performance of CRGAT, which further validates the effectiveness of the proposed CRAGT. The features embedded by the attention mechanism captures the hidden impact of the inter-user interference, which enhances the NNs to learn the optimal mapping from ${\bf H}$ to ${\bf W}$ thus ensuring that the output ${\bf W}$ is almost always in the feasible domain.

Compared with traditional convex optimization based approaches, the inference time is greatly reduced by the learning-based approaches. Besides, the increment number of antennas and users has limited impact on the inference time of the learning-based approaches while greatly increasing the inference time of the convex optimization based approaches. That is, with limited optimality performance loss, the proposed CRGAT is able to make a real-time response to the time-varying CSI, e.g., several microseconds with our computer configuration. The following comparisons are among the learning-based approaches, and the inference time is omitted.

Furthermore, we evaluate the robustness of the proposed CRGAT with $(N_{\rm T},K)=(16,8)$ by adding channel estimation errors to the original input, i.e, $\widehat{\textbf H} = {\textbf{H}} + {\textbf{E}}$ ($\forall k\in\mathcal{K}$), where ${\textbf{H}}$ represents the original input and ${\textbf{E}}$ denotes the CSI error with ${\textbf{E}}_{(k,:)}\sim\mathcal{C} \mathcal{N}(0,\sigma^{2}_{{\rm E},k}{\bf I}_{N_{\rm T}})$ ($\forall k\in\mathcal{K}$). The numerical results are shown in Table \ref{table:10}. It is observed that the adding noise to the original input has limited impact on the optimality performance of the CRGAT.
\begin{table}[ht]
\centering
\caption{The robustness of the CRGAT  with imperfect CSI.}
\begin{tabular}{c|c|c}
\hline
$\sigma^{2}_{{\rm E},k}$ & OP & FR \\
 \hline
$1 \times10^{-4}\|{\textbf{H}}_{(k,:)}\|^2$&{95.3\%}& {100\%}\\
  \hline
$4 \times 10^{-4}\|{\textbf{H}}_{(k,:)}\|^2$&{95.2\%}& {100\%}\\
  \hline
$9 \times 10^{-4}\|{\textbf{H}}_{(k,:)}\|^2$&{95.0\%}& {100\%}\\
 \hline
\end{tabular}
\label{table:10}
\end{table}

\subsection{Scalability}

\subsubsection{Scalability to different numbers of users}
Due to the dynamic nature of the network typology, the generalization capability to the number of users is of high importance. Therefore, this subsection first evaluates the scalability of the well-trained CRGAT, where the numbers of users in the training set and the test set are different. For comparison, the CGCN, CGAT and CTGCN are also simulated, where the CMLP is not included as it is not scalable due to its fixed input ports. The numerical results are shown in Table \ref{table:6}.

%An important test of the usefulness of neural network design is that it can be generalized to different scenarios. In this subsection, we will further evaluate the generalization capability of CRGAT to reveal its adaptability to dynamic network scenarios. The scenarios that can be extended here are specifically in terms of the number of users and the constraints on ${P_{\rm Max}}$.
%\subsubsection{Scale to different users}
%Specifically, after CRGAT is trained in a specific scenario, it can be directly applied in different user scenarios with the same number of antennas.

\begin{table}[ht]
\centering
\caption{Scalability to different numbers of users.}
\begin{tabular}{c|c|c|c|c|c}
\hline
$N_{\rm T}$ &$K_{\rm Tr}$&$K_{\rm Te}$&Model& OP & FR\\
 \hline
 \multirow{8}*{8}&\multirow{8}*{4}&\multirow{4}*{3}& CGCN &82.7\%&100\% \\
\cline{4-6}
 ~&~&~& CGAT &86.4\%&100\% \\
\cline{4-6}
 ~&~&~& CTGCN &89.0\%&100\% \\
\cline{4-6}
 ~&~&~& {\bf CRGAT} &96.0\%&100\% \\
\cline{3-6}
 ~&~&\multirow{4}*{5}& CGCN &78.8\%&92.7\% \\
\cline{4-6}
 ~&~&~& CGAT &82.6\%&98.2\% \\
\cline{4-6}
 ~&~&~& CTGCN &77.8\%&94.3\% \\
\cline{4-6}
  ~&~&~&{\bf CRGAT} &92.2\%&99.5\% \\
\hline
\multirow{8}*{16}&\multirow{8}*{8}&\multirow{4}*{7}& CGCN &55.9\%&99.8\% \\
\cline{4-6}
 ~&~&~& CGAT &84.8\%&100\% \\
\cline{4-6}
 ~&~&~& CTGCN &47.8\%&98.3\% \\
\cline{4-6}
 ~&~&~& {\bf CRGAT} &95.3\%&100\% \\
\cline{3-6}
 ~&~&\multirow{4}*{9}& CGCN &52.2\%&95.9\% \\
\cline{4-6}
 ~&~&~& CGAT &82.5\%&99.6\% \\
\cline{4-6}
 ~&~&~& CTGCN &45.0\%&92.7\% \\
\cline{4-6}
  ~&~&~&{\bf CRGAT} &94.6\%&100\% \\
\hline
\end{tabular}
 \begin{tablenotes}
        \footnotesize
       \item $K_{\rm Tr}$: The number of users in the training set.
       \item $K_{\rm Te}$: The number of users in the test set.
\end{tablenotes}
\label{table:6}
\end{table}
 It is observed that the CRGAT trained on the dataset with $K_{\rm Tr}=4$ (Dataset No. 2) can also achieve a reasonably good performance (within $10\%$ performance loss) when tested on the dataset with  ${K_{\rm Te}\in\{3,5\}}$ (Dataset No. 1 and 3). Besides, the performance loss is within $6\%$ when testing the CRGAT trained on the dataset with $K_{\rm Tr}=4$ (Dataset No. 5) by the dataset with ${K_{\rm Te}\in\{7,9\}}$ (Dataset No. 6 and 7). The reason that the CRGAT yields a better scalability with latter case is that in the former case ($K_{\rm Tr}=4$), $K_{\rm Te}$ is increased/decreased by $25\%$ of $K_{\rm Tr}$ while in the latter case ($K_{\rm Tr}=8$), $K_{\rm Te}$ is increased/decreased by $12.5\%$ of $K_{\rm Tr}$. In addition, the CRGAT outperforms other GNN-based models in terms of scalability. The reason is also due to the attention-enabled aggregation and the residual-assisted combination. Such a result indicates that the proposed CRGAT is scalable to dynamic network typologies without re-training.

\subsubsection{Scalability to different power budgets}
Expect the number of users, the generalization capability of the proposed approach with regard to ${P_{\rm Max}}$ is also evaluated. Specifically, considering a scenario of $(N_{\rm T},K,R_{\rm Req})=(16,8,1)$, we train the CRGAT with $P_{\rm Max}=1$ (Dataset No. 6) while testing the CRGAT with $P_{\rm Max}=2$ (Dataset No. 7) and $P_{\rm Max}=3$ (Dataset No. 8), respectively. For comparison, the following two benchmark methods are also simulated:
\begin{itemize}
  \item [1)] 
  {\bf PM only (PM-only) \cite{PM1}:} All constraints are handled by the PM.   
  \item [2)]
  {\bf LDM only (LDM-only) \cite{LDM}:} All constraints are handled by the LDM.   
\end{itemize}
The numerical results are shown in Table \ref{table:9}. It is observed all methods under test achieve $100\%$ feasibility rate, while the proposed methods are able to obtain higher optimality performance than the benchmarking ones. Note that the generalization capability of the proposed approach can save time and arithmetic power to re-train the neural network for different scenarios.

\begin{table}[ht]
\centering
\caption{Scalability to different $P_{\rm Max}$.}
\begin{tabular}{c|c|c|c|c}
\hline
${P_{\rm Max}^{\rm Tr}}$&${P_{\rm Max}^{\rm Te}}$&Method & PER & FR \\
 \hline
\multirow{8}*{1}&\multirow{4}*{2}& PM-only &78.5\%&100\% \\
\cline{3-5}
~&~& LDM-only &78.3\%&100\% \\
\cline{3-5}
~& ~& {\bf PM with (\ref{af})} &92.4\%&100\% \\
\cline{3-5}
~&~& {\bf LDM with (\ref{af})} &92.3\%&100\% \\
\cline{2-5}
~&\multirow{4}*{3}& PM-only &71.1\%&100\% \\
\cline{3-5}
~&~& LDM-only &70.1\%&100\% \\
\cline{3-5}
 ~&~& {\bf PM with (\ref{af})} &90.1\%&100\% \\
\cline{3-5}
~&~& {\bf LDM with (\ref{af})} &89.9\%&100\% \\
 \hline
\end{tabular}
 \begin{tablenotes}
        \footnotesize
       \item ${P_{\rm Max}^{\rm Tr}}$: $P_{\rm Max}$ in the training set. 
       \item ${P_{\rm Max}^{\rm Te}}$: $P_{\rm Max}$ in the test set. 
\end{tablenotes}
\label{table:9}
\end{table}

\subsection{Adaptability}

The adaptability is to evaluate the performance of the learning-based approaches under different scenarios. The adaptability to different $R_{\rm Req}$ and ${P_{\rm Max}}$ is respectively evaluated as follows.

%In our proposed approach, the constraints are handled by the proposed activation function (\ref{af}) and the loss functions (based on PM and LDM), which are evaluated in this subsection with $(N_{\rm T},K)=(16,8)$.

Table \ref{table:7} shows the results with different values of $R_{\rm Req}$. It is observed that with an increase of $R_{\rm Req}$, the optimality performance of each method almost keeps unchanged while the feasible rate decreases in general. The reason is that with larger value of $R_{\rm Req}$, the feasible set of problem (\ref{pb}) shrinks. Besides, the proposed methods are superior to the benchmark ones, especially in terms of the feasible rate. As it is of high significance for learning-based approaches to yield feasible solutions, the effectiveness of the proposed approach is validated. 
\begin{table}[ht]
\centering
\caption{Adaptability to different ${R_{\rm Req}}$.}
\begin{tabular}{c|c|c|c}
\hline
${R_{\rm Req}}$&Method & OP & FR\\
 \hline
\multirow{4}*{1}& PM-only &94.5\%&63.0\% \\
\cline{2-4}
~& LDM-only &94.5\%&84.7\% \\
\cline{2-4}
 ~&  {\bf PM with (\ref{af})} &95.4\%&100\% \\
\cline{2-4}
~&  {\bf LDM with (\ref{af})} &95.2\%&100\% \\
 \hline
\multirow{4}*{2}& PM-only &94.3\%&62.0\% \\
\cline{2-4}
~& LDM-only &94.5\%& 91.3\% \\
\cline{2-4}
 ~& {\bf PM with (\ref{af})} &94.9\%&99.6\% \\
\cline{2-4}
~& {\bf LDM with (\ref{af})} &95.6\%&99.7\% \\
 \hline
\multirow{4}*{3}& PM-only &94.4\%&51.0\% \\
\cline{2-4}
~& LDM-only &95.1\%&85.3\% \\
\cline{2-4}
 ~& {\bf PM with (\ref{af})} &94.8\%&94.1\% \\
\cline{2-4}
~& {\bf LDM with (\ref{af})} &95.4\%&94.3\% \\
 \hline
\end{tabular}
\label{table:7}
\end{table}

Table \ref{table:8} shows the results with different values of ${P_{\rm Max}}$. It is interesting to see that with an increment of ${P_{\rm Max}}$, the optimality performance of all methods is slightly degraded. This might be due to the fact that with larger ${P_{\rm Max}}$, there are more allocation schemes of the power budget, and  the learning task becomes more complicated. Nevertheless, the feasibility rates of the proposed methods keeps $100\%$, which outperforms the benchmark methods.
\begin{table}[ht]
\centering
\caption{Adaptability to different ${P_{\rm Max}}$.}
\begin{tabular}{c|c|c|c}
\hline
${P_{\rm Max}}$& Method & OP & FR \\
 \hline
\multirow{4}*{1}& PM-only &94.5\%&63.0\% \\
\cline{2-4}
~& LDM-only &94.5\%&84.7\% \\
\cline{2-4}
 ~& {\bf PM with (\ref{af})} &95.4\%&100\% \\
\cline{2-4}
~& {\bf LDM with (\ref{af})} &95.2\%&100\% \\
 \hline
\multirow{4}*{2}& PM-only &91.3\%&71.4\% \\
\cline{2-4}
~& LDM-only &91.5\%&88.4\% \\
\cline{2-4}
 ~& {\bf PM with (\ref{af})} &92.6\%&100\% \\
\cline{2-4}
~& {\bf LDM with (\ref{af})} &92.1\%&100\% \\
 \hline
\multirow{4}*{3}& PM-only &88.5\%&74.5\% \\
\cline{2-4}
~& LDM-only &89.9\%&87.3\% \\
\cline{2-4}
 ~& {\bf PM with (\ref{af})} &89.7\%&100\% \\
\cline{2-4}
~& {\bf LDM with (\ref{af})} &90.5\%&100\% \\
 \hline
\end{tabular}
\label{table:8}
\end{table}

%\subsubsection{Robustness to imperfect inputs}

\subsection{Ablation experiment: Residual-assisted combination}

The effectiveness of the attention-enabled aggregation has been analyzed in our previous work \cite{li_gat}, which is also validated in Table \ref{table:5} and Table \ref{table:6}. For a deeper understanding of the residual-assisted combination, this subsection shows the performance of CGAT and CRGAT with different depths. The numerical results are shown in Table \ref{table:3}.

\begin{table}[ht]
\centering
\caption{Ablation experiment: Residual-assisted combination.}
\begin{tabular}{c|c|c|c}
\hline
 Model &Depth& OP & FR \\
 \hline
CGAT & 2 &84.4\% & 99.9\%\\
 \hline
 CGAT & 3 &84.0\% & 99.9\%\\
 \hline
 CGAT & 4 &4.0\%$^*$ & 0\%\\
 \hline
 CRGAT & 2 &84.9\% & 100\%\\
 \hline
 CRGAT & 3 &93.8\% & 100\%\\
 \hline
 CRGAT & 4 &95.4\% & 100\%\\
 \hline
\end{tabular}
 \begin{tablenotes}
        \footnotesize
       \item Depth: Total number of complex graph convolution layers.
\end{tablenotes}
\label{table:3}
\end{table}

It is observed that with the increment of depths, the performance of the CGAT is degraded. In particular, a sharp decay is induced with $4$ depths. This is due to the oversmoothing issue as analyzed in Remark \ref{Rem1}. However, the oversmoothing issue is avoided by the CRGAT, as the expression capacity of the CRGAT is enhanced with the increment of depths. Such a result validates the effectiveness and the importance of the residual-assisted combination.

Furthermore, we adopt the mean average distance (MAD) \cite{MAD} among the node features to quantify oversmoothing issue, which is calculated by
\begin{flalign}
 {\rm MAD} = \frac{\sum_{i=1}^K\sum_{j=1}^K \mathbf{D}_{(i,j)}}{\sum_{i=1}^K \left|\mathcal{N}\left(i\right)\right| },
\end{flalign}
where $\mathbf{D} \in{\mathbb{R}^{{K}\times{K}}}$ denotes the distance matrix which is calculated by
\begin{flalign}
\mathbf{D}_{(i,j)} = 1 - \frac{\overline{\mathbf{H}}^{(l)}_{(i,:)} \overline{\mathbf{H}}^{(l)}_{(j,:)}}{\Vert \overline{\mathbf{H}}^{(l)}_{(i,:)} \Vert_2 \Vert \overline{\mathbf{H}}^{(l)}_{(j,:)} \Vert_2},
\end{flalign}
where $\overline{\mathbf{H}}^{(l)} =  {\rm Com}\left({\rm Re}(\mathbf{H}^{(l)}),{\rm Im}(\mathbf{H}^{(l)}) \right) \in{\mathbb{C}^{{K}\times{2{{\widetilde F}(l)}}}}$.
The MADs of different layers in CGAT and CRGAT are shown in Table \ref{table:4}.

\begin{table}[ht]
\centering
\caption{MAD in different layers of CGAT and CRGAT.}
\begin{tabular}{c|c|c|c|c}
\hline
\diagbox{Model}{Layer}  &1& 2 &3&4 \\
 \hline
CGAT & 0.382 & 0.303 & 0.238 & 0.086\\
 \hline
 CRGAT & 0.712 &0.689  & 0.484& 0.472\\
 \hline
\end{tabular}
\label{table:4}
\end{table}

%Formally, given the graph representation matrix $\mathbf{H} \in{\mathbb{C}^{{K}\times{{{\widetilde F}(l)}}}}$ , we first obtain the distance matrix $\mathbf{D} \in{\mathbb{R}^{{K}\times{K}}}$ for $\mathbf{H}$ by computing the cosine distance between each node pair:
It is observed that the MAD of the CGAT converges to $0$ with a growth of the depths, which indicates that the node features intend to be indistinguishable. However, a larger MAD is obtained in the CRGAT. As the learning capability is roughly positive correlation to the depths, the residual-assisted combination in fact balances the trade-off between enhancing the learning capability and avoiding the oversmoothing issue.

\subsection{Transfer learning}

As analysed before, the scalability to different numbers of users of the CRGAT is effected by $|K_{\rm Tr}-K_{\rm Te}|/K_{\rm Tr}$. To improve the optimality performance, instead of generalizing, one can also reuse of the pre-trained CRGAT on a new test set in a transfer learning manner. 

Fig. \ref{transfer} show the transfer learning performance of the CRGAT, where the learnable parameters of the CRGAT trained on the dataset with $K_{\rm Tr}=4$ (Dataset No. 2)  is leveraged as the initial parameters for fine-tuning the CRGAT on the dataset with $K_{\rm Tr}=5$ (Dataset No. 3). For comparison, the convergence behavior of retraining the CRGAT on the dataset with $K_{\rm Tr}=5$ (Dataset No. 3) is also plotted. It is observed that with the transfer learning, the CRGAT is able to achieve about $95\%$ optimality performance after $4$ epochs while the generalization performance is about $92\%$ (in Table \ref{table:6}). Besides, the convergence speed of the transfer learning is much faster than that of the retraining with a slight optimality performance gain. 

%The generalized digestion of CRGAT, as shown in Table \ref{table:6}, is affected by the degree of variation in the number of users in the environment, and it can be observed that the performance of CRGAT trained on the dataset with $K_{\rm Tr}=4$ ($N_{\rm T}=8$) on $K_{\rm Te}=5$ is only about 90\%. However, the parameters already trained by CRGAT make sense as initial parameters for transfer learning, and we use the parameters trained on $K =4$ as initial parameters for fine-tuning on the $K=5$ dataset. The convergence results of transfer learning and training from new are shown in Fig.\ref{transfer}. It can be seen that the convergence speed of transfer learning is much faster than the latter, which avoids the waste of computational resources.

\begin{figure}[ht]
{\centering
{\includegraphics[ width=.48\textwidth]{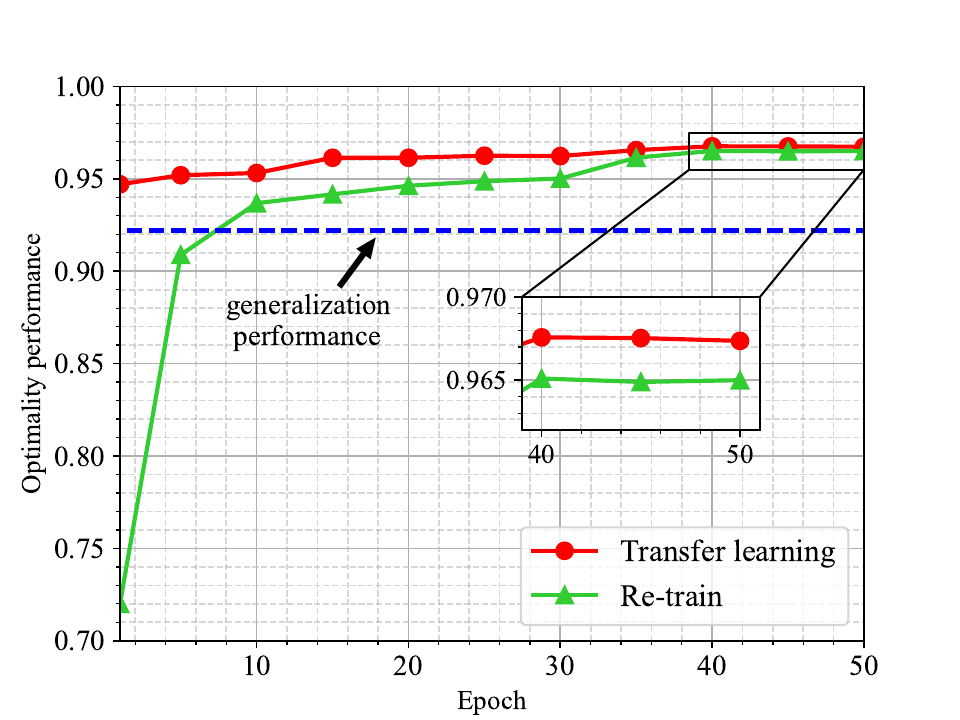}}}
\caption{{Convergence behavior of the transfer learning and re-train.}}
\label{transfer}
\end{figure}

\section{CONCLUSION}

In this paper, we studied an end-to-end GNN-based beamforming design for a constrained sum-rate maximization problem. A GNN-based architecture named CRGAT was proposed to directly map the CSI to the beamforming vectors. To enhance the model representation and avoid the oversmoothing issue, the attention mechanisms and residual structures were applied to the message passing mechanism. To guarantee the feasible solution, a novel activation function was proposed to handle the sum-power constraint, and two loss functions based were designed for the QoS requirements. The CRGAT was trained in an unsupervised learning manner, and was evaluated from the perspectives of model effectiveness, scalability and adaptability.  Compared with the bechmarking schemes, the advantages of the proposed CRGAT are summarized as: 1) real-time response to time-varying CSI with limited optimality performance loss; 2) scalability to dynamic network typologies; 3) adaptability to various network settings.

\begin{appendices}
\section{SCA-based Solution Approach for the problem (\ref{pa})}

The problem (\ref{pa}) can be solved by the SCA method, and the details are given as follows.

By defining auxiliary variables $\{\gamma_k\}$ satisfying that $\gamma_k \le {R_{k}}( {\{ {{{\bf{w}}_{i}}} \}})$, the problem (\ref{pa}) is equivalent to 
\begin{subequations}\label{p2}
\begin{align}
&\mathop {\max }\limits_{\left\{{{{\bf{w}}_{i}}},\gamma_i \right\}} { {\sum\nolimits_{k = 1}^K \gamma_k  } }  \label{cons:p2:A}\\
{\rm s.t.}~&{\left| {{\bf{h}}_{k}^H{{\bf{w}}_{k}}} \right|}^2 \ge \left(2^{\gamma_k}-1\right)\left({{{ \sum\nolimits_{i=1,i\ne k}^K {{\left| {{\bf{h}}_{k}^H{{\bf{w}}_{i}}} \right|}^2} + \sigma _{{k}}^2}}}\right),\label{cons:p2:B}\\
&{ {\sum\nolimits_{k = 1}^K {\left\| {{{\bf{w}}_{k}}} \right\|_2^2} } } \le {P_{\rm Max}},\label{cons:p2:C}\\
&\gamma_k \ge {R_{\rm Req}},\label{cons:p2:D}\\
&{{\bf{w}}_{i}}\in{\mathbb{C}^{{N_{\rm{T}}}}},{\forall i,k}\in{\cal K}.\nonumber
\end{align}
\end{subequations}

By defining the auxiliary variables $\{a_i,b_i\}$ satisfying that
\begin{flalign}
\left\{ \begin{array}{l}
{\left| {{\bf{h}}_k^H{{\bf{w}}_k}} \right|^2} \ge {e^{{a_k} + {b_k}}}\\
{e^{{a_k}}} \ge {2^{{\gamma _k}}} - 1\\
{e^{{b_k}}} \ge \sum\nolimits_{i = 1,i \ne k}^K {{{\left| {{\bf{h}}_k^H{{\bf{w}}_i}} \right|}^2}}  + \sigma _k^2
\end{array} \right.,
\end{flalign}
the problem (\ref{p2}) is equivalent to 
\begin{subequations}\label{p3}
\begin{align}
&\mathop {\max }\limits_{\left\{{{{\bf{w}}_{i}}},\gamma_i,a_i,b_i \right\}} { {\sum\nolimits_{k = 1}^K \gamma_k  } }  \label{cons:p3:A}\\
{\rm s.t.}~&{\left| {{\bf{h}}_k^H{{\bf{w}}_k}} \right|^2} \ge {e^{{a_k} + {b_k}}},\label{cons:p3:B}\\
&{e^{{a_k}}} \ge {2^{{\gamma _k}}} - 1,\label{cons:p3:C}\\
&{e^{{b_k}}} \ge \sum\nolimits_{i = 1,i \ne k}^K {{{\left| {{\bf{h}}_k^H{{\bf{w}}_i}} \right|}^2}}  + \sigma _k^2,\label{cons:p3:D}\\
&{\rm (\ref{cons:p2:C}),(\ref{cons:p2:D})},{{\bf{w}}_{i}}\in{\mathbb{C}^{{N_{\rm{T}}}}},{\forall i,k}\in{\cal K}.\nonumber
\end{align}
\end{subequations}

Apply the first order Taylor approximation to non-convex constraints (\ref{cons:p3:B}), (\ref{cons:p3:C}) and (\ref{cons:p3:D}), the problem (\ref{p3}) is approximated by the following convex problem (\ref{p4}) with given feasible $\{{{{\widetilde{\bf w}}}_i},{{\widetilde a}_i},{{\widetilde b}_i}\}$:
\begin{subequations}\label{p4}
\begin{align}
&\mathop {\max }\limits_{\left\{{{{\bf{w}}_{i}}},\gamma_i,a_i,b_i \right\}} { {\sum\nolimits_{k = 1}^K \gamma_k  } }  \label{cons:p4:A}\\
{\rm s.t.}~&2{\mathop{\rm Re}\nolimits} \left\{ {{{{{{{\widetilde{\bf w}}}_k}}}^H}{{\bf h}}_k{{\bf h}}_k^H {{{\bf{w}}_k}}} \right\} - {\left| {{{\bf h}}_k^H{{{\widetilde{\bf w}}}_k}} \right|^2} \ge {e^{{a_k} + {b_k}}},\\
&e^{{\widetilde a}_k}\left(1+a_k - {\widetilde a}_k \right)\ge 2^{\gamma_k}-1,\\
&e^{{\widetilde b}_k}\left(1+b_k - {\widetilde b}_k \right)\ge \sum\nolimits_{i=1,i\ne k}^K {{\left| {{\bf{h}}_{k}^H{{\bf{w}}_{i}}} \right|}^2}+\sigma _{{k}}^2,\\
&{\rm (\ref{cons:p2:C}),(\ref{cons:p2:D})},{{\bf{w}}_{i}}\in{\mathbb{C}^{{N_{\rm{T}}}}},{\forall i,k}\in{\cal K}.\nonumber
\end{align}
\end{subequations}

By iteratively solving the problem (\ref{p4}) via the SCA method, a near-optimal solution to the problem (\ref{pa}) can be obtained. Specially, the initialization $\{{{\widetilde{\bf w}}}_i\}$ can be obtained by solving the following second order cone programming. 
\begin{subequations}
\begin{align}
 {\rm Find } &~ {{\left\{ {{{\bf{w}}_i}} \right\}} }  \nonumber\\
{\rm s.t.}&~\sqrt{\left(1+\frac{1}{2^{R_{\rm Req}}-1}\right)}{{{{{ {\bf h}}_k^H{{\bf{w}}_k}} }}}\ge \left\| {\bm \Omega }_k{\bf f} + {\bf b}_k \right\|,\\
&{\rm Re}\left\{{{ {\bf h}}_k^H{{\bf{w}}_k}} \right\}\ge0, {\rm Im}\left\{{{ {\bf h}}_k^H{{\bf{w}}_k}} \right\}=0,\\
&{\rm (\ref{cons:p2:C})},{{\bf{w}}_{i}}\in{\mathbb{C}^{{N_{\rm{T}}}}},{\forall i,k}\in{\cal K}.\nonumber
\end{align}
\end{subequations}
where
\begin{flalign}\nonumber
\left\{ \begin{array}{l}
{\bf{f}} \triangleq {\left[ {{\bf{w}}_1^H,{\bf{w}}_2^H,...,{\bf{w}}_K^H} \right]^H}\in{\mathbb C}^{KN_{\rm T}}\\
{{\bf{b}}_k} \triangleq {\left[ {{\bf{0}}_K^T,\sigma _k^2} \right]^T}{ \in {\mathbb C}^{K+1}}\\
{\Omega _k} \triangleq \left[ {\begin{array}{*{20}{l}}
{{\rm{diag}}\left( {{{\bf{h}}_k}^H,{{\bf{h}}_k}^H,...,{{\bf{h}}_k}^H} \right)}\\
{{\bf{0}}_{K{N_{\rm{T}}}}^T}
\end{array}} \right]\in{\mathbb C}^{(K+1)\times {KN_{\rm T}}}
\end{array} \right..
\end{flalign}
With initialized $\{{{\widetilde{\bf w}}}_i\}$, ${{\widetilde a}_i}$ and ${{\widetilde b}_i}$ are respectively initialized by 
\begin{flalign}\nonumber
\left\{ \begin{array}{l}
{e^{{a_k}}} = {2^{{R_{k}}( {\{ {{{\widetilde{\bf{w}}}_{i}}} \}} )}} - 1\\
{e^{{b_k}}} = \sum\nolimits_{i = 1,i \ne k}^K {{{\left| {{\bf{h}}_k^H{{{\widetilde{\bf{w}}}_{i}}}} \right|}^2}}  + \sigma _k^2
\end{array} \right..
\end{flalign}

\section{MRT and ZF beamforming design}

In the MRT and ZF beamforming designs, each beamforming vectors are simplified by fixing its direction as 
\begin{flalign}
{\bf w}_k={\sqrt p_k}{{\overline{\bf w}}_k},
\end{flalign}
where $p_k\in{\mathbb R}^+$ and ${{\overline{\bf w}}_k}$ satisfying $\|{{\overline{\bf w}}_k}\|_2^2=1$ denote the power component and the direction component, respectively.

In the MRT design, the direction of beamforming is set as ${{\bf h}_k}/{\|{\bf h}_k\|_2}$, and in the ZF design the direction of beamforming is set as ${{\bf v}_k}/{\|{\bf v}_k\|_2}$, where ${{\bf v}_k}$ is the $k$-th column of ${\bf V}\triangleq {\bf G}^H({\bf G}{\bf G}^H)^{-1}$ and ${\bf G}\triangleq[{\bf h}_1^H,{\bf h}_2^H,...,{\bf h}_K^H]$.

The power components in the MRT and ZF designs can also be optimized by the SCA method, which is similar to the SCA-based solution approach for the problem (\ref{pa}), and thus, omitted here. 

%Besides, the optimal MRT and ZF designs can learned by the proposed GAT-based approach, where the activation function to meet the sum-power constraint is given by
%\begin{flalign}
%\phi_2\left( p_k  \right) = \left\{ \begin{array}{l}
%\sqrt {{p_k^{\left({\rm C}\right)}}{P_{{\rm{Max}}}}} ,\sum\nolimits_{{\rm{i}} = 1}^K {{p_i^{\left({\rm C}\right)}}}  \le 1\\
%\sqrt {\frac{{{p_k^{\left({\rm C}\right)}}}}{{\sum\nolimits_{{\rm{i}} = 1}^K {{p_i^{\left({\rm C}\right)}}} }}{P_{{\rm{Max}}}}} ,\sum\nolimits_{{\rm{i}} = 1}^K {{p_i^{\left({\rm C}\right)}}}  > 1
%\end{array} \right.
%\end{flalign}
%where $p_k^{\left({\rm C}\right)}$ is the output of the $C$-th CFCL.

\end{appendices}

\end{document}